\documentclass{pic2012}
\def\runauthor{P. Van Mulders on behalf of the ATLAS and CMS Collaborations}
\def\shorttitle{Searches for new fermions and new bosons}

\begin{document}

\title{Searches for new fermions and new bosons}

\author{P. Van Mulders \\
On behalf of the ATLAS and CMS collaborations
}

\address{E-mail: pvmulder@cern.ch }

\maketitle

\abstracts{
The high center-of-mass energy at the LHC provides the opportunity to test the predictions of some of the beyond the standard model theories. We provide an overview of a selected number of searches for new fermions and new bosons with the ATLAS and CMS experiments. No evidence for the existence of new particles was found and therefore limits are obtained on the parameters of the models under consideration.
}

\section{Introduction} 
The standard model of particle physics is a very succesful theory describing most of our current experimental knowledge. There are however a number of observations that are not explained, such as for example the existence of dark matter and dark energy, the nonzero masses of the neutrinos and the baryon asymmetry. Many exotic models exist that extend the standard model and provide an answer to some of the open questions. New models typically imply the existence of new particles at a higher energy scale. Some of the models can be probed at the energy scales that are currently reached at the LHC. Many searches for new particles have been performed by the ATLAS~\cite{ATLAS} and CMS~\cite{CMS} experiments using the 7 TeV proton collisions that are collected in 2011 and they continued these searches using the 8 TeV data. Since no evidence is found for the existence of new particles, the searches result in upper limits on the production cross section of the new processes, excluding parts of the allowed parameter space. 

Given the wide variety of exotic models and searches, we focus on the most recent and the most stringent results from both experiments. In this document we provide an overview of the searches for new fermions and new bosons, while searches for other exotic phenomena such as for instance extra dimensions, long-lived particles and leptoquarks are presented in Ref.~\cite{Santanastasio}. 
%Section~\ref{sect:Z'} summarizes the searches for high-mass resonances decaying to two oppositely charged leptons or two jets. The results are interpreted in the context of models predicting the existence of a heavy $Z'$ boson. For some analyses an interpretation is also given in the context of other theories. Searches for a new charged boson, denoted with $W'$, are presented in Section~\ref{sect:W'}. Apart from new bosons, also new fermions are predicted by many theories. Some of those predict for instance the existence of heavy neutrinos. Limits on the existence of these neutrinos are provided in Section~\ref{sect:N}. Section~\ref{sect:TB} overviews some of the recent searches for a fourth generation of quarks. Finally, a grand overview is given in Section~\ref{sect:summary}.

\section{Searches for new $Z$ bosons}
\label{sect:Z'}
Several beyond the standard model theories predict the existence of a new neutral gauge boson $Z'$. Depending on the model, the $Z'$ boson may decay either into leptons or jets, leading to a resonance in the reconstructed dilepton or dijet mass distributions respectively. Different searches are performed depending on the reconstructed objects in the final state: two oppositely charged electrons or muons, two tau leptons or two jets. The resulting limits on the production cross section are interpreted in benchmark models, such as the sequential standard model (SSM) that predicts a $Z'$ boson with the same couplings to fermions as the standard model $Z$ boson.

\begin{figure}[!thb]
  \centering
  \includegraphics[width=0.49 \textwidth]{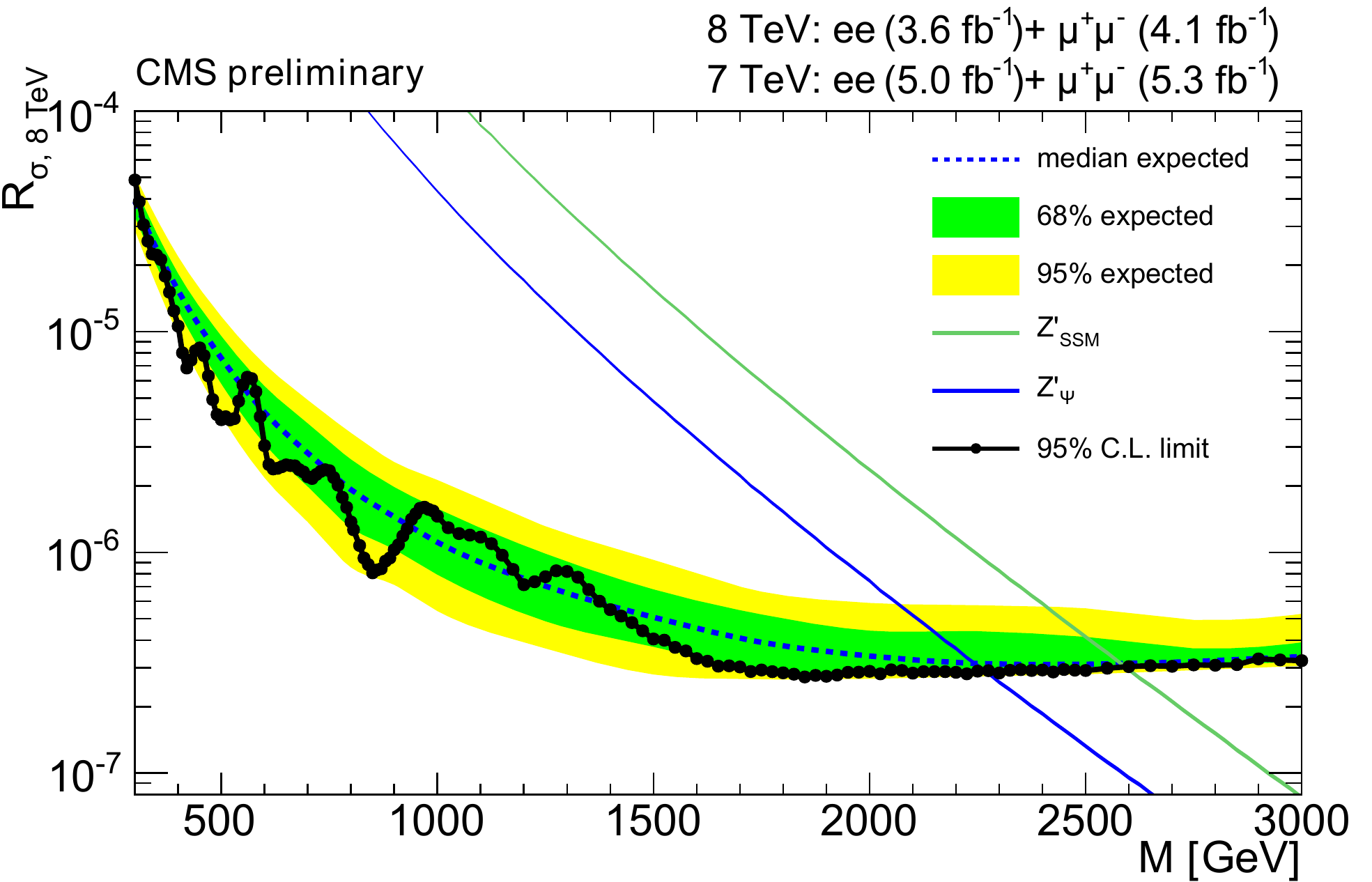}
  \includegraphics[width=0.49 \textwidth]{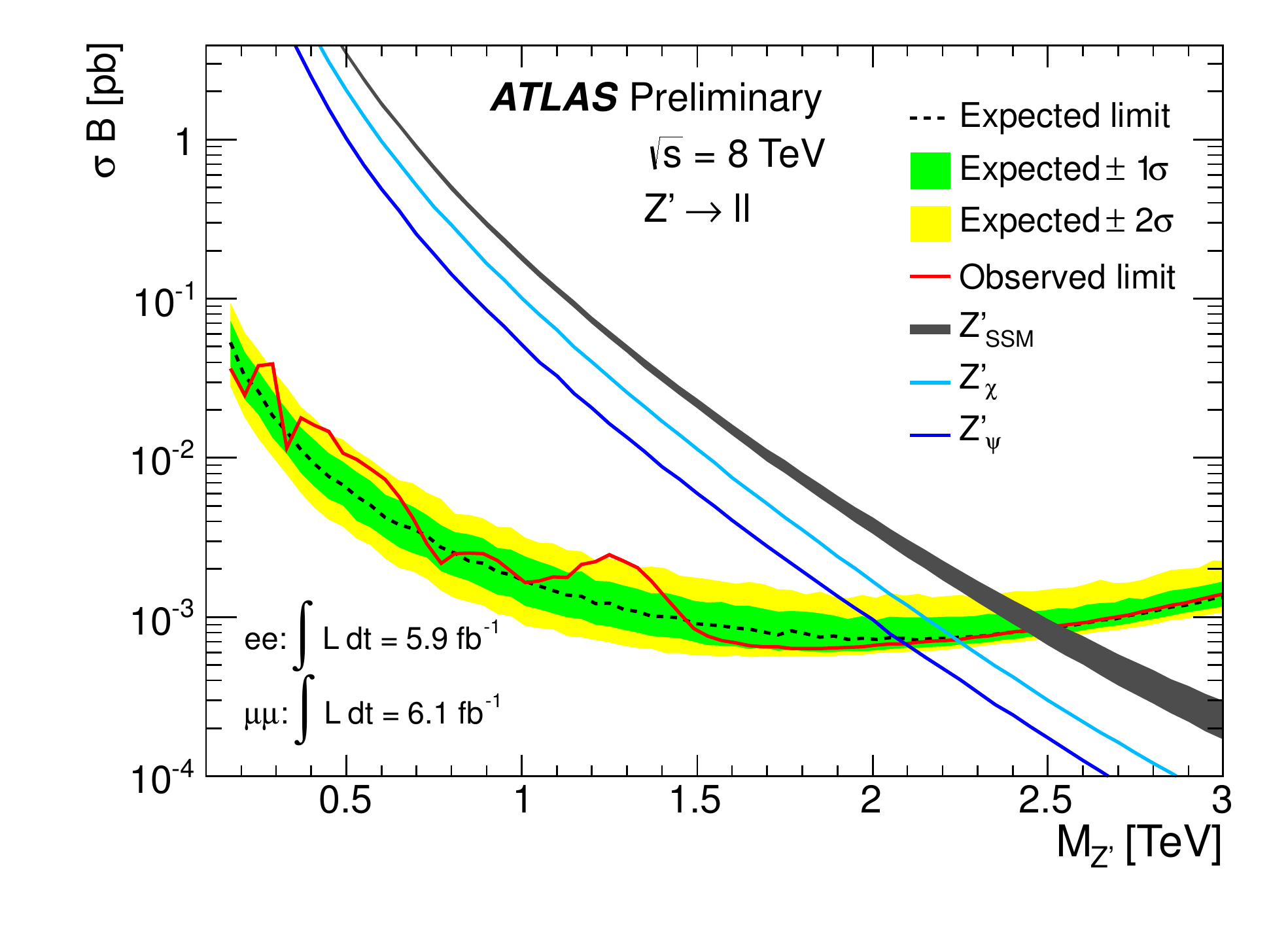}
\caption[*]{The upper limit on the ratio of the production cross section times branching ratio of the signal $Z'$ boson and the production cross section times branching ratio of the standard model $Z$ boson as a function of the mass of the $Z'$ boson for the CMS experiment (left) and the upper limit on the production cross section times branching ratio as a function of the mass of the $Z'$ boson for the ATLAS experiment (right).}
  \label{fig:Z'll}
\end{figure}

If the $Z'$ boson decays to two oppositely charged electrons or muons, the invariant mass of the dielectron or dimuon pair is reconstructed for each selected event. The resulting dielectron and dimuon invariant mass distributions are then fitted for the presence of a resonance. In the absence of any evidence for a new resonance upper limits are determined on the production cross section times branching ratio of the $Z'$ boson to dileptons. Both the CMS and ATLAS collaborations have performed this search on the 7 and 8 TeV data sets~\cite{CMSZ'll,ATLASZ'll}. Figure~\ref{fig:Z'll} shows the resulting limits. The CMS (ATLAS) experiment excludes the existence of a SSM $Z'$ boson with a mass below 2.59 (2.49) TeV at the 95\% confidence level (CL). It is worth noting that the result from the CMS experiment is obtained from the combination of the 7 TeV (5 fb$^{-1}$) and 8 TeV (4 fb$^{-1}$) data sets, while the result of the ATLAS experiment is obtained using only the 8 TeV data set (6 fb$^{-1}$).  

\begin{figure}[!thb]
  \centering
  \includegraphics[width=0.49 \textwidth]{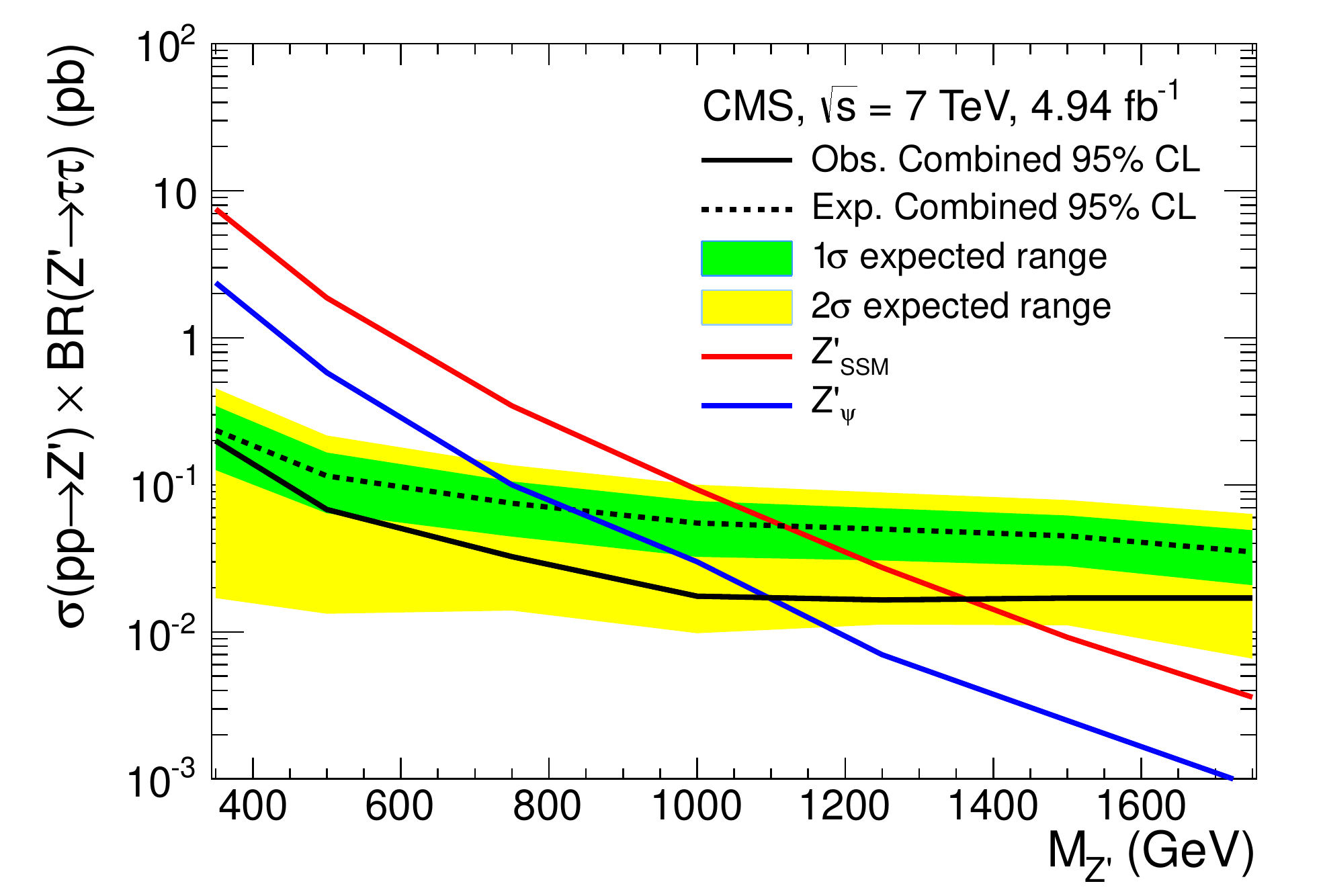}
  \includegraphics[width=0.49 \textwidth]{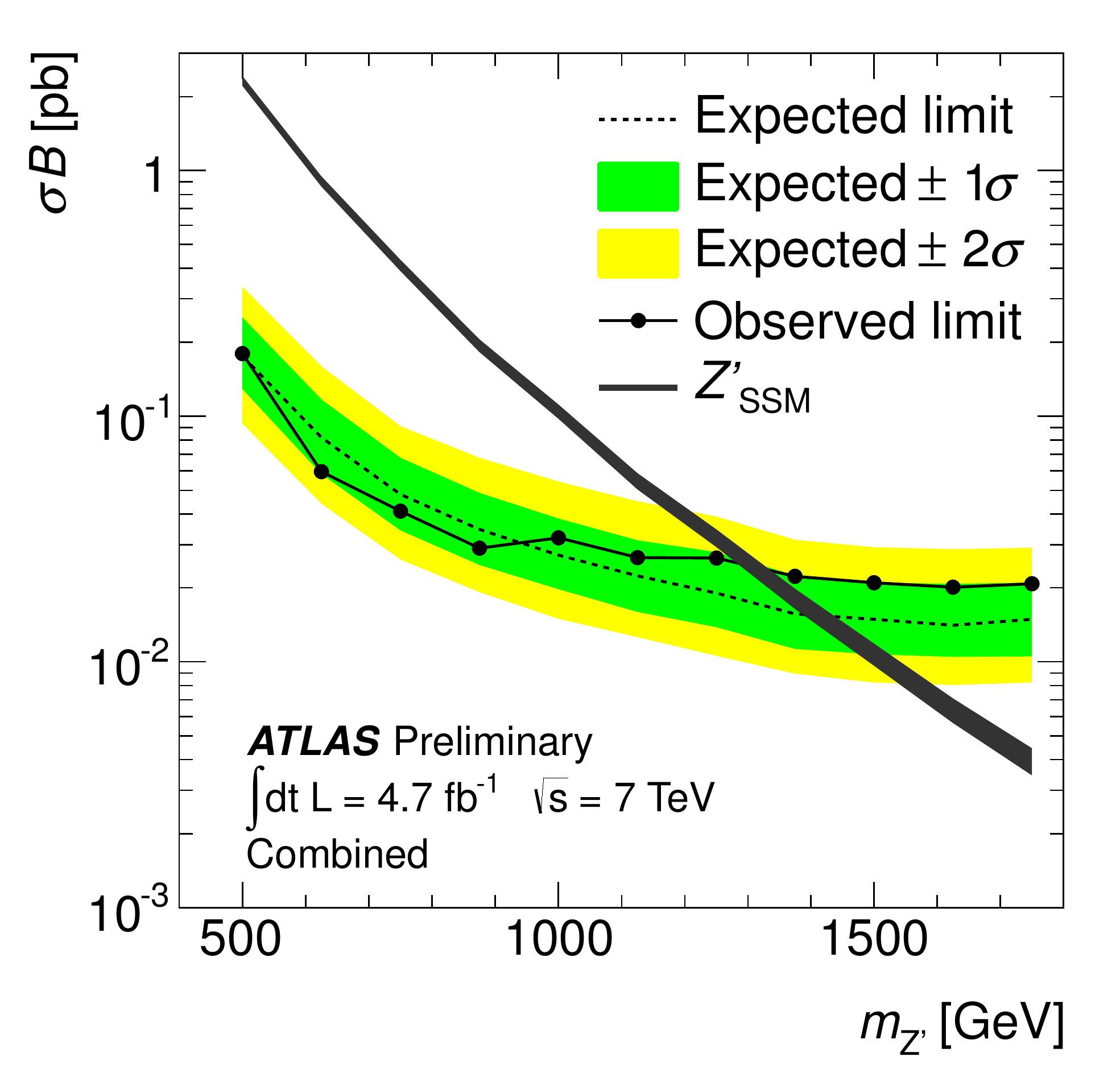}
\caption[*]{The upper limit on the production cross section times branching ratio as a function of the mass of the $Z'$ boson for the CMS (left) and ATLAS (right) experiments.}
  \label{fig:Z'tautau}
\end{figure}

In case the $Z'$ boson would decay to two oppositely charged tau leptons, the analysis becomes more complicated. The tau lepton may decay either to an electron or muon and two neutrinos, or to charged hadrons, neutrinos and possibly some additional neutral hadrons. Due to the presence of the neutrinos, it is more difficult to reconstruct the invariant mass of the ditau pair. Therefore, the CMS collaboration reconstructs the effective visible mass $M(\tau_1,\tau_2,E_{t}^{miss})=\sqrt{(E_{\tau_1}+E_{\tau_2}+E_{t}^{miss})^2 - (\vec{p_{\tau_1}}+\vec{p_{\tau_2}}+\vec{E_{t}^{miss}})^2}$. The effective visible mass distribution is then fitted to obtain the upper limit on the production cross section times branching ratio of the $Z'$ boson to ditaus~\cite{CMSZ'tautau}. The ATLAS collaboration uses the transverse mass {\small{$M_T=\sqrt{2p_{T\tau_1}p_{T\tau_2}(1-\mathrm{cos}\Delta\phi_{\tau_1,\tau_2})+2\not\!\!E_{t}p_{T\tau_1}(1-\mathrm{cos}\Delta\phi_{\tau_1,miss})+2\not\!\!E_{t}p_{T\tau_2}(1-\mathrm{cos}\Delta\phi_{\tau_2,miss})}$}} as a sensitive observable, where $\not\!\!\!E_{t}$ denotes the missing transverse momentum. Upper limits are obtained by counting the number of events with $M_T$ above a certain threshold value, that is optimized for each signal mass~\cite{ATLASZ'tautau}. The resulting limits are shown in Figure~\ref{fig:Z'tautau}. The CMS (ATLAS) experiment excludes the existence of a SSM $Z'$ boson with a mass below 1.4 (1.3) TeV at the 95\% CL using the full 2011 data set which corresponds to an integrated luminosity of about 5 fb$^{-1}$.

\begin{figure}[!thb]
  \centering
  \includegraphics[width=0.49 \textwidth]{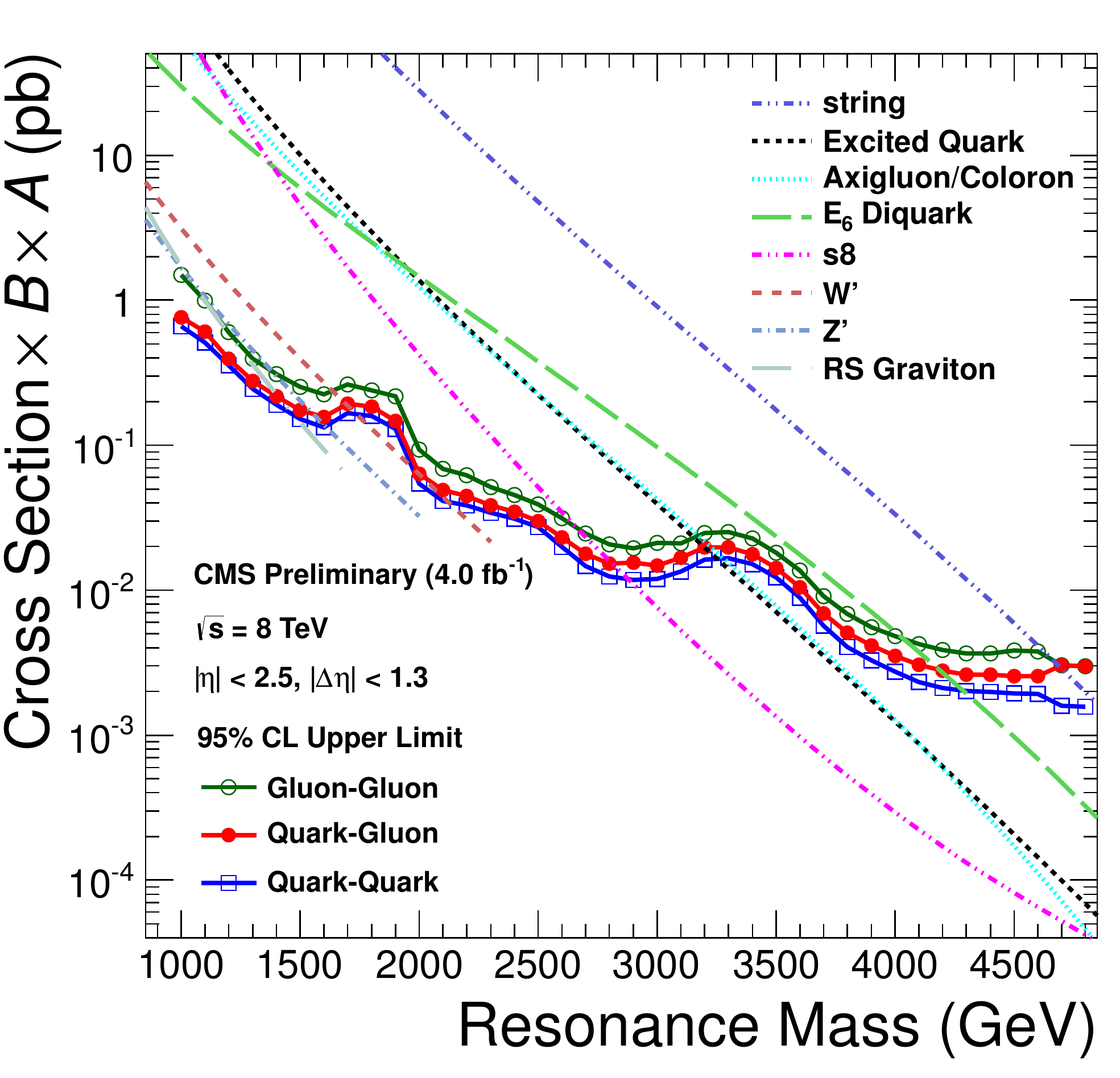}
  \includegraphics[width=0.49 \textwidth]{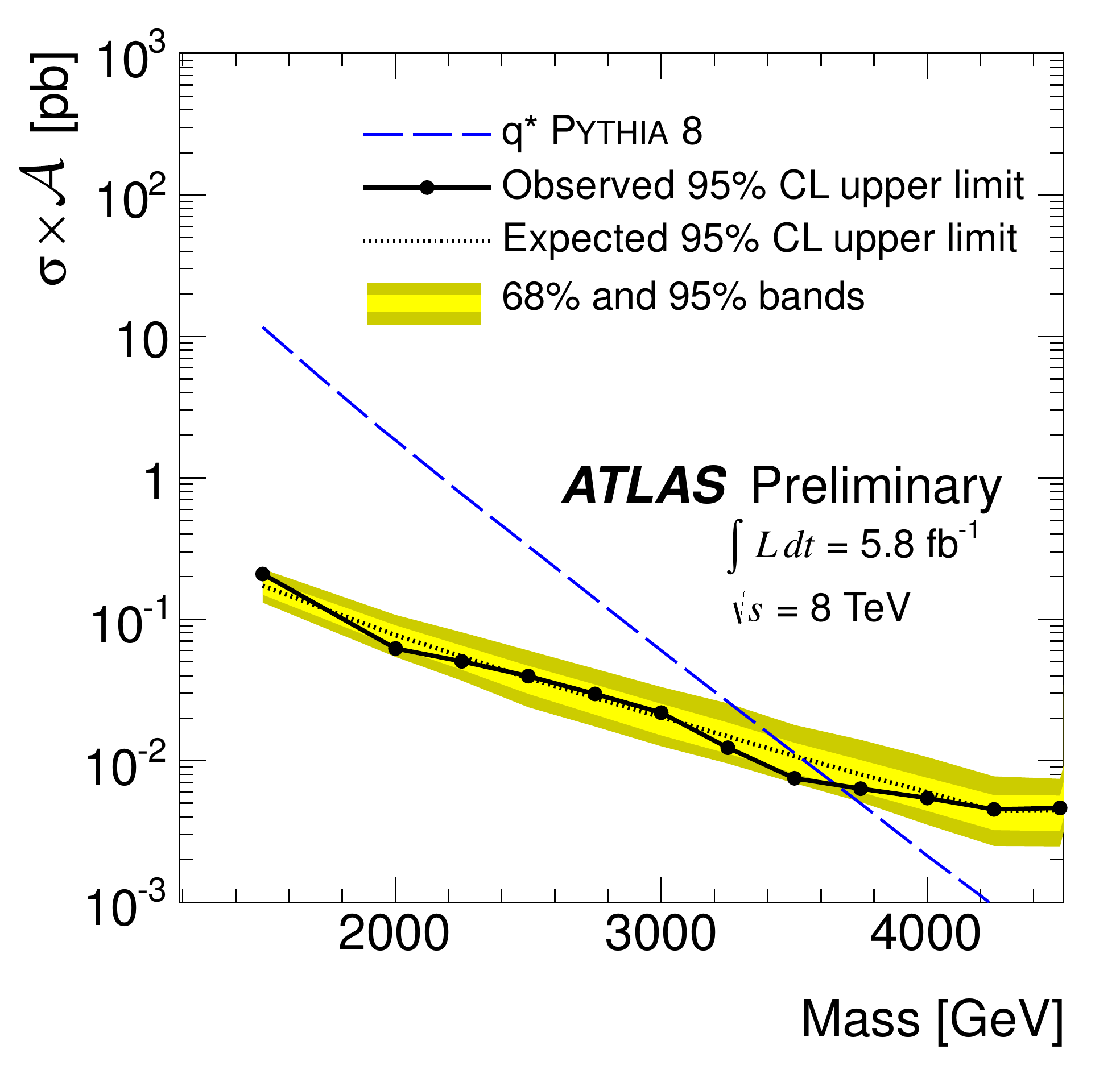}
\caption[*]{The upper limit on the production cross section times branching ratio as a function of the mass of the heavy resonance for the CMS (left) and ATLAS (right) experiments.}
  \label{fig:Z'qq}
\end{figure}

Some models predict a $Z'$ or $W'$ boson that decays to two quarks. Therefore, searches are developed which require two well-separated jets with a high transverse momentum~\cite{CMSZ'dijet,ATLASZ'dijet}. The CMS experiment reconstructs `wide' jets by taking the two jets with the highest transverse momentum and merge them with other jets that are nearby. From these two jets, the dijet invariant mass distribution is reconstructed and fitted with a smooth functional form: $f(x)=p_1(1-x)^{p_2}x^{p_3+p_4\mathrm{ln}x}$, with $x\equiv m_{jj}/\sqrt{s}$. The resulting upper limit on the production cross section times branching ratio is shown in Figure~\ref{fig:Z'qq}. The result of the ATLAS experiment is interpreted in a model that predicts an excited quark decaying to two jets. The same model is also shown for CMS experiment. The ATLAS (CMS) collaboration excludes the existence of an excited quark with a mass below 3.8 (3.2) TeV at the 95\% CL. In addition, the CMS experiment also excludes the existence of a SSM $Z'$ boson with a mass below 1.6 TeV at the 95\% CL.

The CMS collaboration also developed a dedicated search for $b\bar{b}$ resonances~\cite{CMSZ'dibjet}. The efficiency to correctly identify a jet as a $b$ jet decreases with the mass of the heavy resonance. Events with two `wide' jets are selected and categorized in one of the three different subsamples according to the number of jets that are identified as a $b$ jet. In each of these subsamples the dijet invariant mass distribution is reconstructed. The upper limit using the three subsamples is obtained for different values of the branching fraction to b quarks. For the SSM $Z'$ boson, this branching fraction is about 0.22, and its existence is excluded with a mass below 1.5 TeV at the 95\% CL. If a model predicts a larger branching fraction to b quarks, the upper limit improves by up to 70\%, excluding a larger part of the allowed mass range.

\section{Searches for new $W$ bosons}
\label{sect:W'}

Charged gauge bosons $W'$ are predicted by various theories beyond the standard model. The simplest model is again the SSM that predicts a $W'$ boson with the same couplings to the fermions as the standard model $W$ boson. Therefore, the $W'$ boson decays either to leptons or to two quarks.
The search for a $W'$ boson decaying to two quarks is covered by the search for dijet resonances discussed in the previous section. In this section we will study the decay of the $W'$ or $W^*$ boson into leptons. Another final state that is considered is the decay of the $W'$ boson into a top quark and another quark, from theories that allow top-flavor-violating processes.

\begin{figure}[!thb]
  \centering
  \includegraphics[width=0.49 \textwidth]{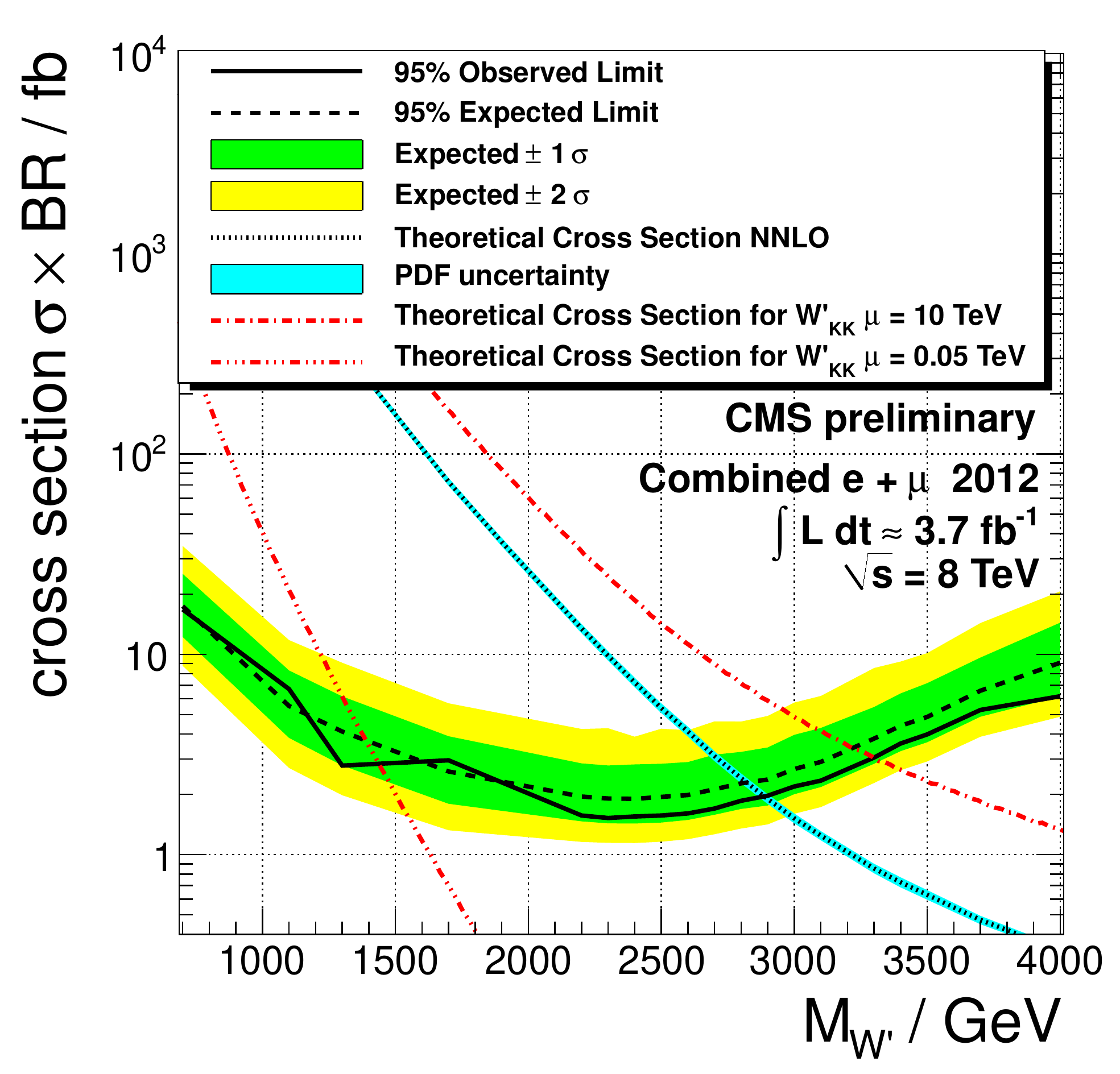}
  \includegraphics[width=0.49 \textwidth]{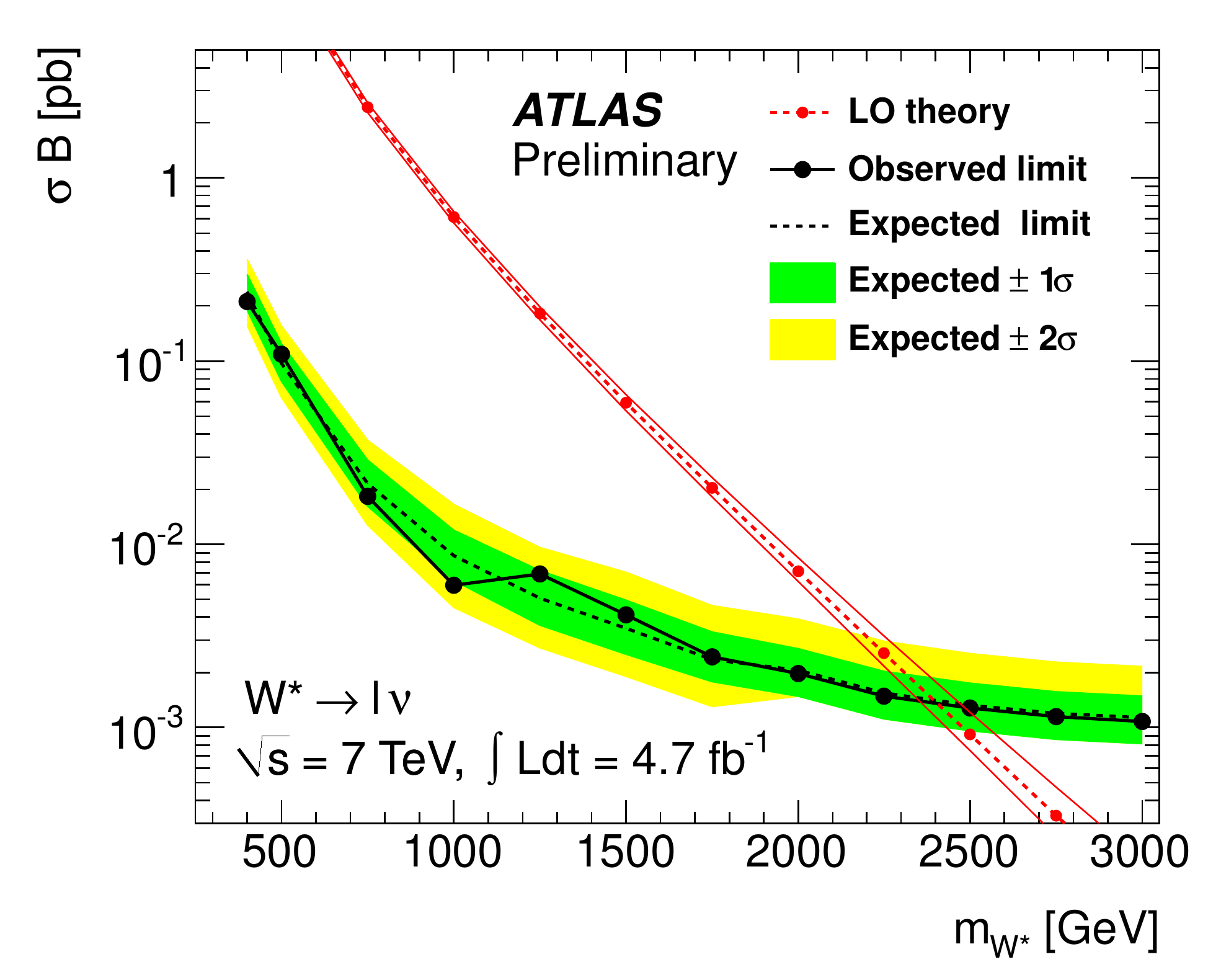}
\caption[*]{The upper limit on the production cross section times branching ratio as a function of the mass of the new $W$ boson for the CMS (left) and ATLAS (right) experiments.}
  \label{fig:W'lv}
\end{figure}

Both the CMS and ATLAS experiments performed a search for a new $W$ boson decaying to a charged lepton and a neutrino~\cite{CMSW'lv,ATLASW'lv}. The transverse mass of the selected events is reconstructed as $M_T=\sqrt{2p_T^lE_T^{miss}(1-\mathrm{cos}\Delta\phi_{l,\nu})}$. The upper limits on the production cross section times branching fraction is obtained from the selected number of events after requiring a minimum transverse mass. The transverse mass threshold is optimized for each signal mass. The resulting upper limit is shown in Figure~\ref{fig:W'lv}. The CMS experiment excludes the existence of a SSM $W'$ boson with a mass below 2.85 TeV at the 95\% CL, while the ATLAS experiment excludes the existence of a $W^*$ with a mass below 2.55 TeV at the 95\% CL.

\begin{figure}[!thb]
  \centering
  \includegraphics[width=0.49 \textwidth]{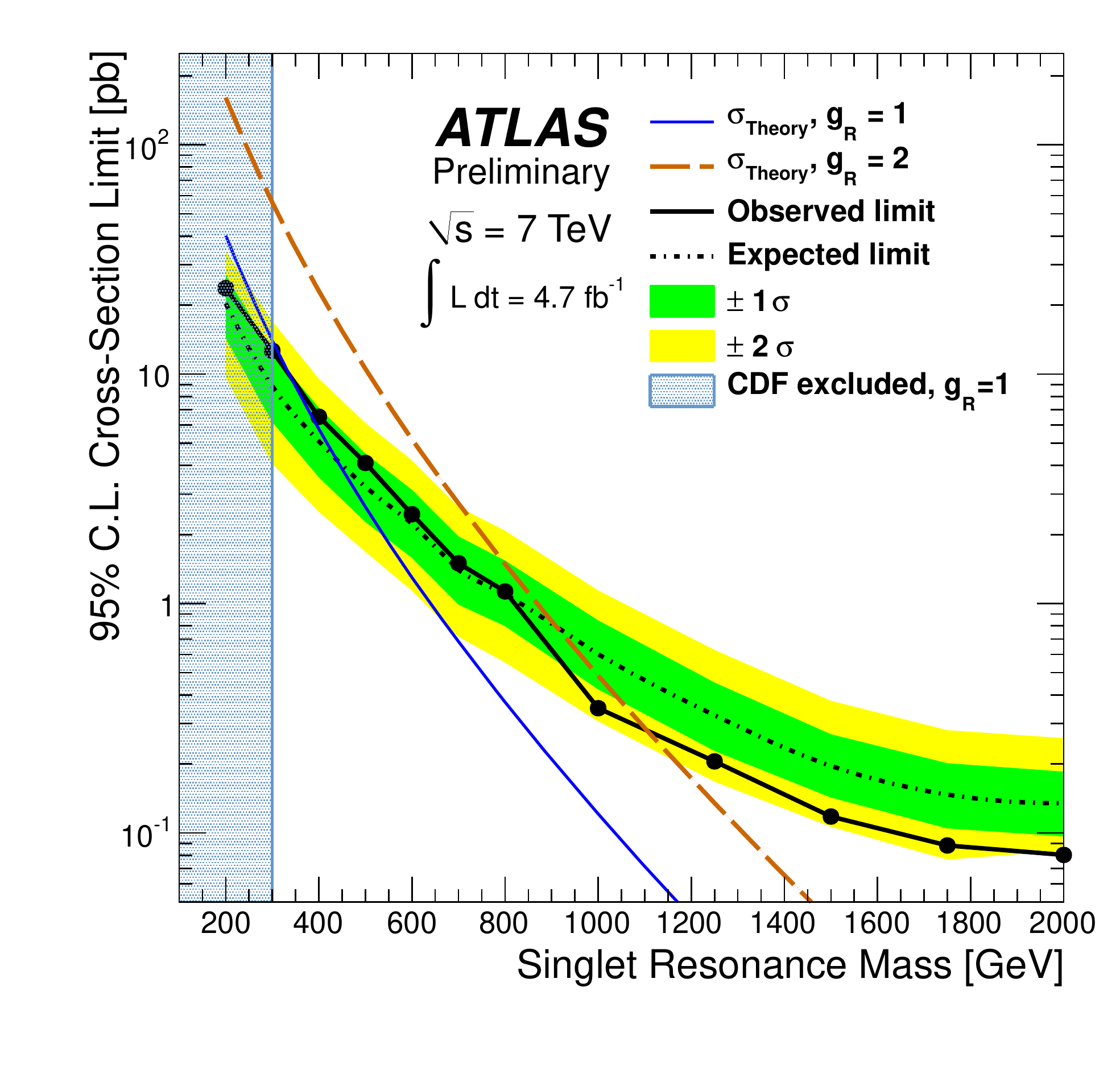}
  \includegraphics[width=0.49 \textwidth]{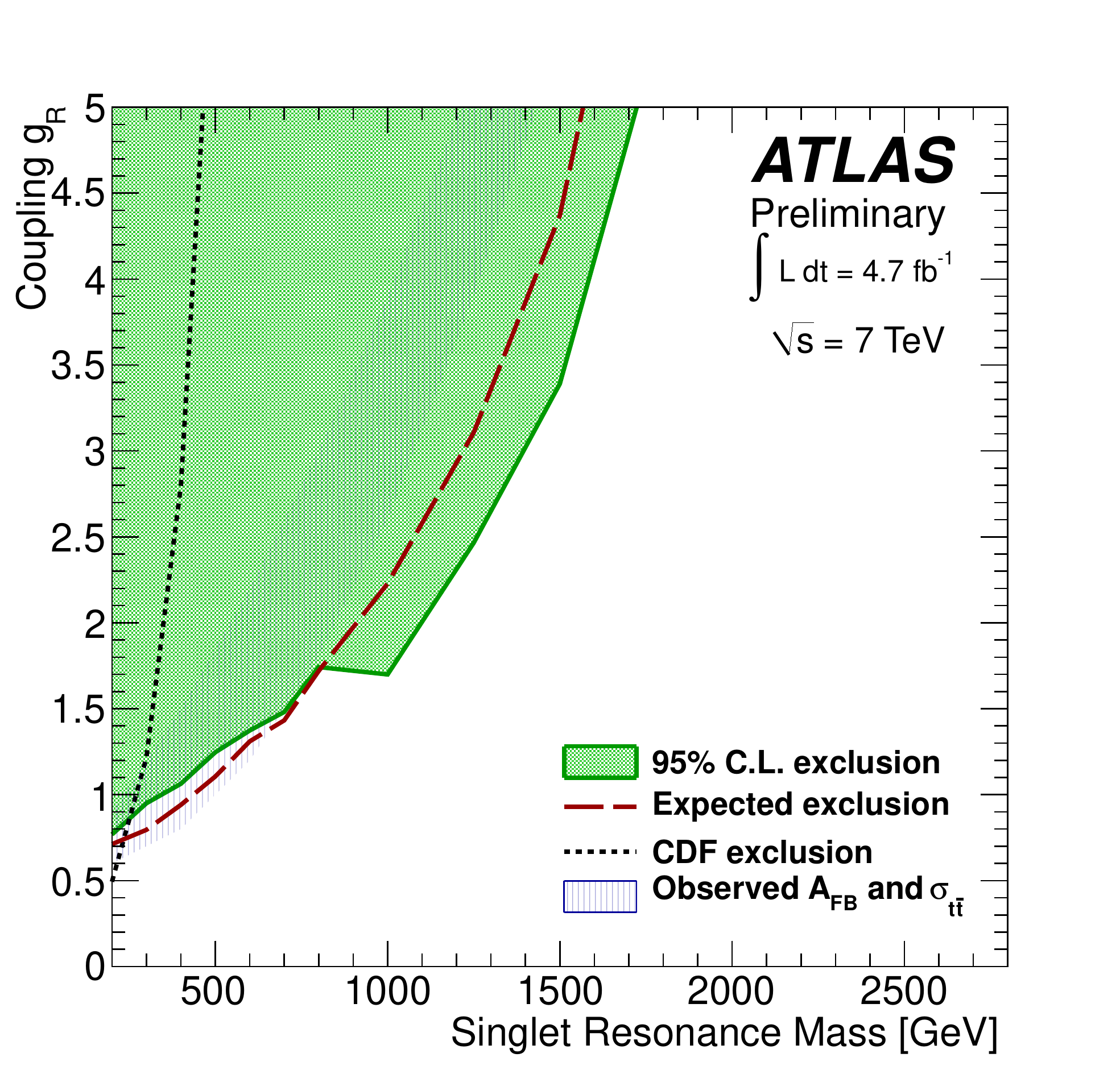}
\caption[*]{The upper limit on the production cross section times branching ratio as a function of the mass of the $W'$ boson (left) and the limit as a function of the coupling parameter $g_R$ and the mass of the $W'$ boson (right).}
  \label{fig:W'tq}
\end{figure}

The forward-backward asymmetry measured in $t\bar{t}$ events at the Tevatron collider is not entirely consistent with the standard model prediction. The discrepancy could be explained by a top-flavor-violating process such as for instance $pp \rightarrow W' t \rightarrow \bar{t}q t$. Either the $W$ boson from the decay of the $t$ or from the decay of the $\bar{t}$ quark is required to decay into leptons, which results in a final state with a charged lepton, missing transverse energy and five jets. Both ATLAS and CMS reconstruct the $W'$ boson mass distributions from the $tq$ or $\bar{t}q$ systems~\cite{CMSW'tq,ATLASW'tq}. Therefore, it is crucial to assign the jets correctly to the quarks from the $W'\rightarrow \bar{t} q$ decay. The ATLAS analysis uses a kinematic likelihood fitter with the $W$ boson and top quark masses as constraints. The remaining jets are paired with the reconstructed top quark and the combinations that yield the largest values of $m_{tq}$ and $m_{\bar{t}q}$ are chosen. In the CMS analysis the jet combination is chosen for which the invariant mass of three jets is closest to the top quark mass. This jet combination is then combined with the jet that has the highest transverse momentum. The upper limit on the production cross section times branching ratio shown in Figure~\ref{fig:W'tq} is obtained by counting the number of events after requiring a minimum value of $m_{tq}$ or $m_{\bar{t}q}$. Only the result for the ATLAS experiment is shown, because of the larger sensitivity. Assuming a right-handed coupling $g_R=2$ the ATLAS (CMS) experiment excludes the existence of a $W'$ boson with a mass below 1.1 TeV (840 GeV) at the 95\% CL. 

\section{Searches for heavy neutrinos}
\label{sect:N}
\begin{figure}[!thb]
  \centering
  \includegraphics[width=0.49 \textwidth]{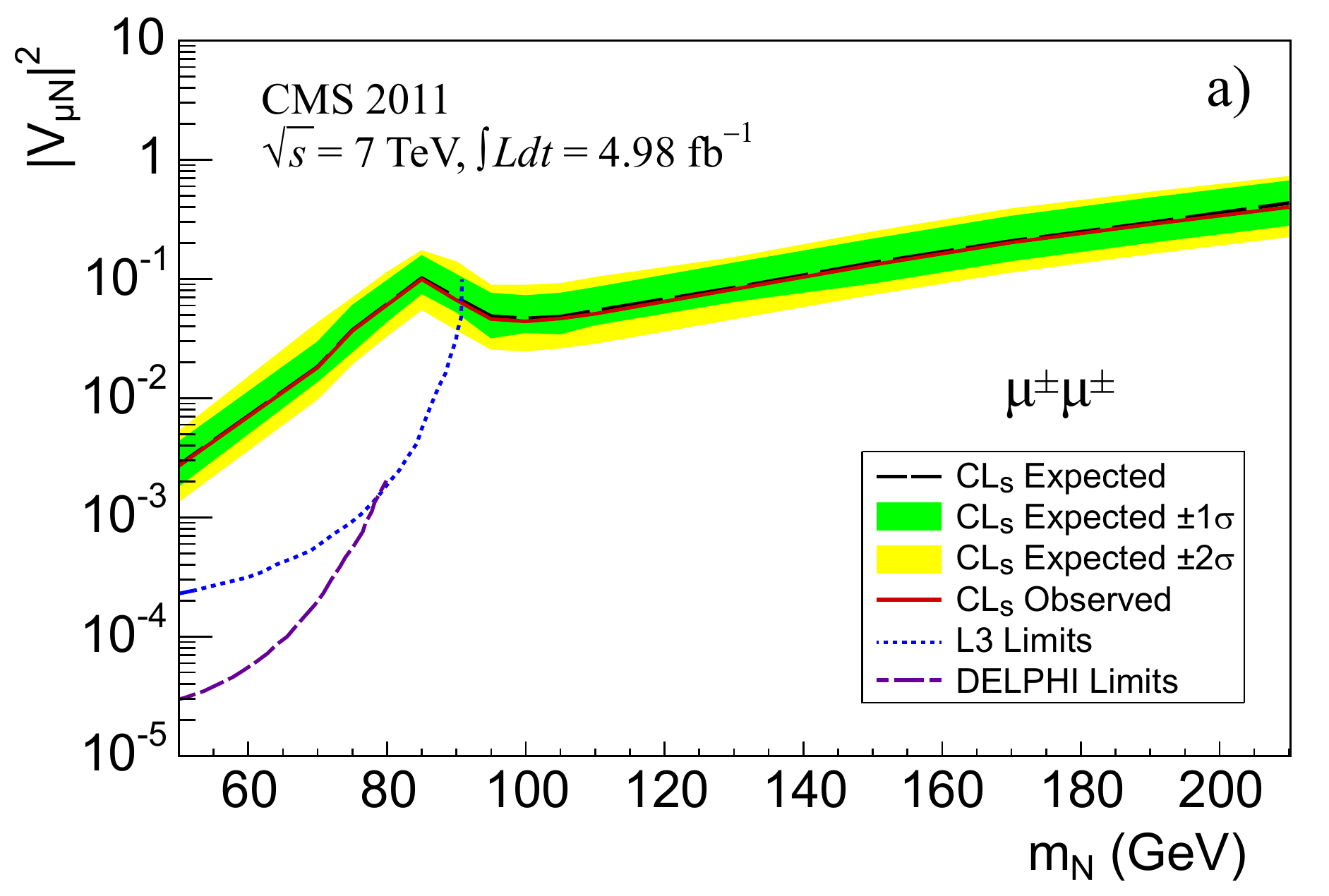}
  \includegraphics[width=0.49 \textwidth]{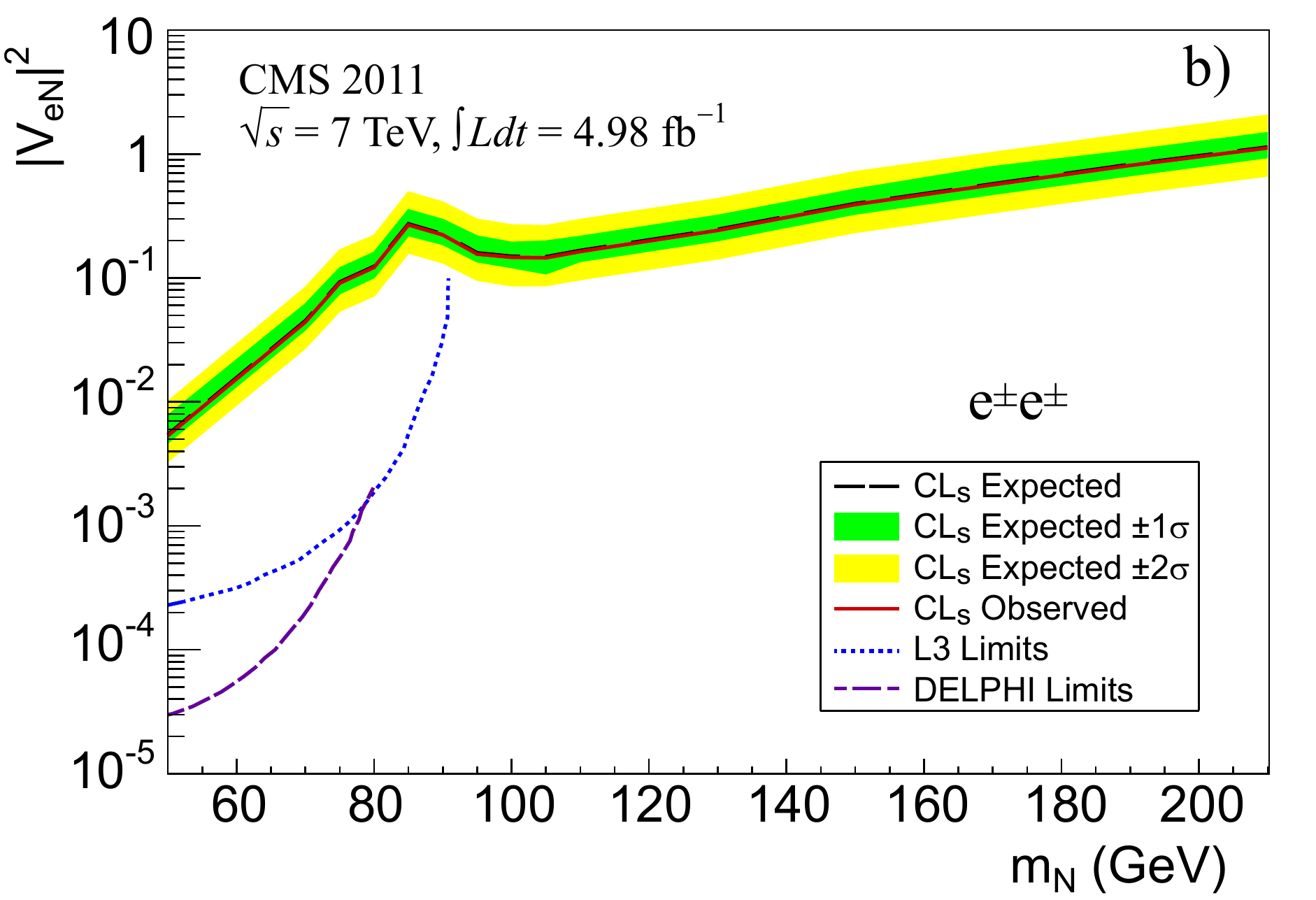}
\caption[*]{The left (right) figure shows the limit on $|V_{\mu N}|^2$ ($|V_{eN}|^2$) as a function of $m_N$ under the assumption $|V_{eN}|^2 = |V_{\tau N}|^2 = 0$ ($|V_{\mu N}|^2 = |V_{\tau N}|^2 = 0$).} 
  \label{fig:NMaj}
\end{figure}

The observation of neutrino oscillations, and consequently nonzero neutrino masses, is a clear indication of physics beyond the standard model. Grand unified theories predict the existence of at least one heavy Majorana neutrino. This heavy Majorana neutrino might then provide masses to the neutrinos through the see-saw mechanism. Majorana neutrinos allow lepton and lepton-flavor number violating interactions. Therefore, searches for Majorana neutrinos are typically performed by looking at final states with two same-sign leptons. The CMS experiment performed a model-independent search for a heavy isosinglet Majorana neutrino~\cite{CMS_Nmaj} with the mass of the heavy Majorana neutrino $m_N$ and the mixing element $V_{lN}$ as free parameters. $V_{lN}$ describes the mixing between the heavy Majorana neutrino and the standard model neutrino of flavor $l$. The considered process is $pp\rightarrow N l^{\pm} \rightarrow l^{\pm} W l^{\pm} \rightarrow l^{pm} q\bar{q} l^{\pm}$. Therefore, the signature consists of two same-sign leptons of the same flavour and at least two jets from the decay of the $W$ boson. Upper limits are set on the square of the mixing element, $|V_{lN}|^2$, with $l=e,\mu$, as a function of the Majorana neutrino mass. The upper limits shown in Figure~\ref{fig:NMaj} are the first direct upper limits on the heavy-Majorana mixing for $m_N>90$ GeV.

For $m_N = 90$ GeV the CMS experiment finds $|V_{\mu N}|^2 < 0.07$ and $|V_eN|^2 < 0.22$. At $m_N = 210$ GeV the limit becomes $|V_{\mu N}|^2 < 0.43$, while for $|V_{eN}|^2$ the limit reaches 1.0 at a mass of 203 GeV.

\begin{figure}[!thb]
  \centering
  \includegraphics[width=0.49 \textwidth]{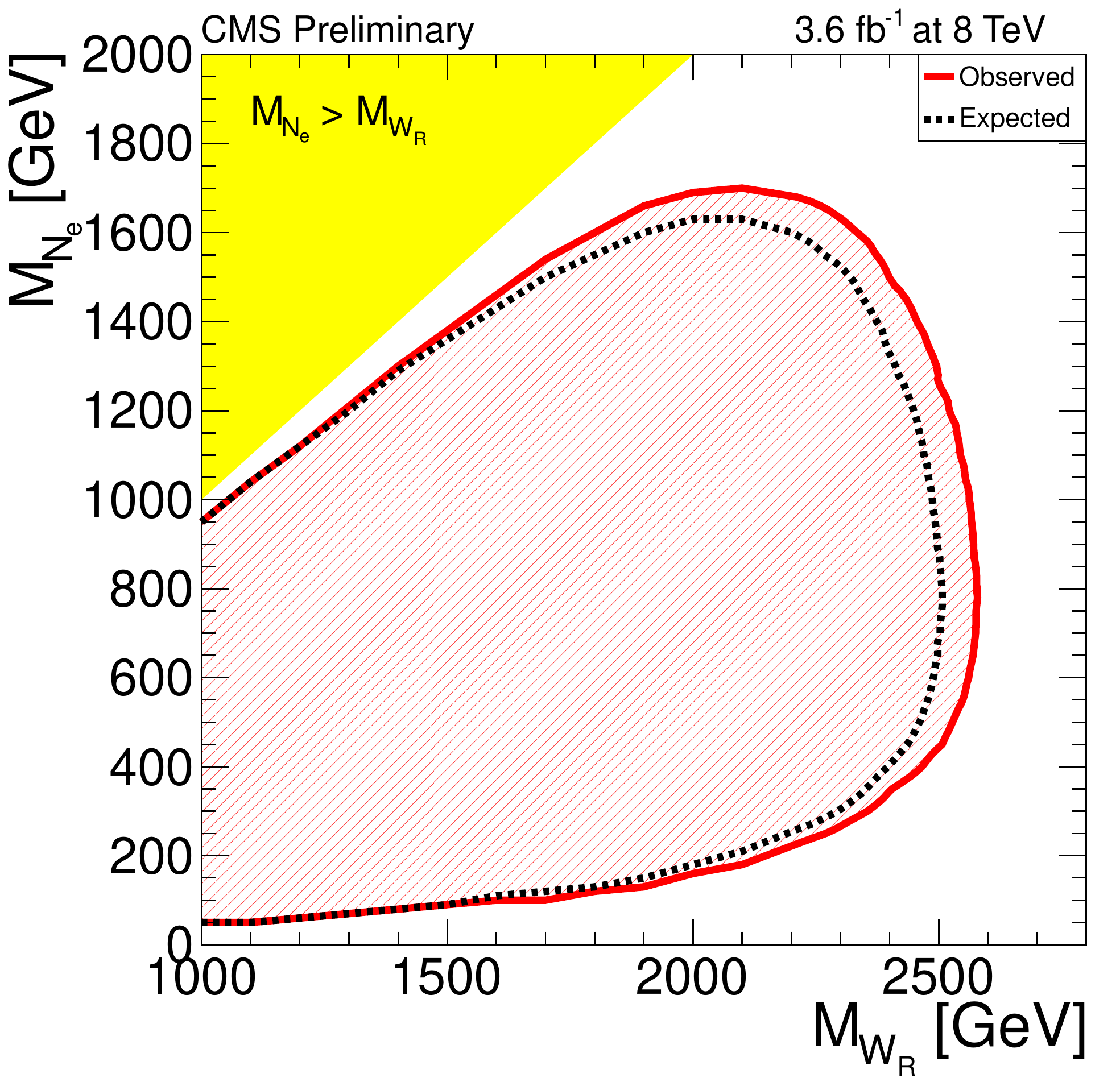}
  \includegraphics[width=0.49 \textwidth]{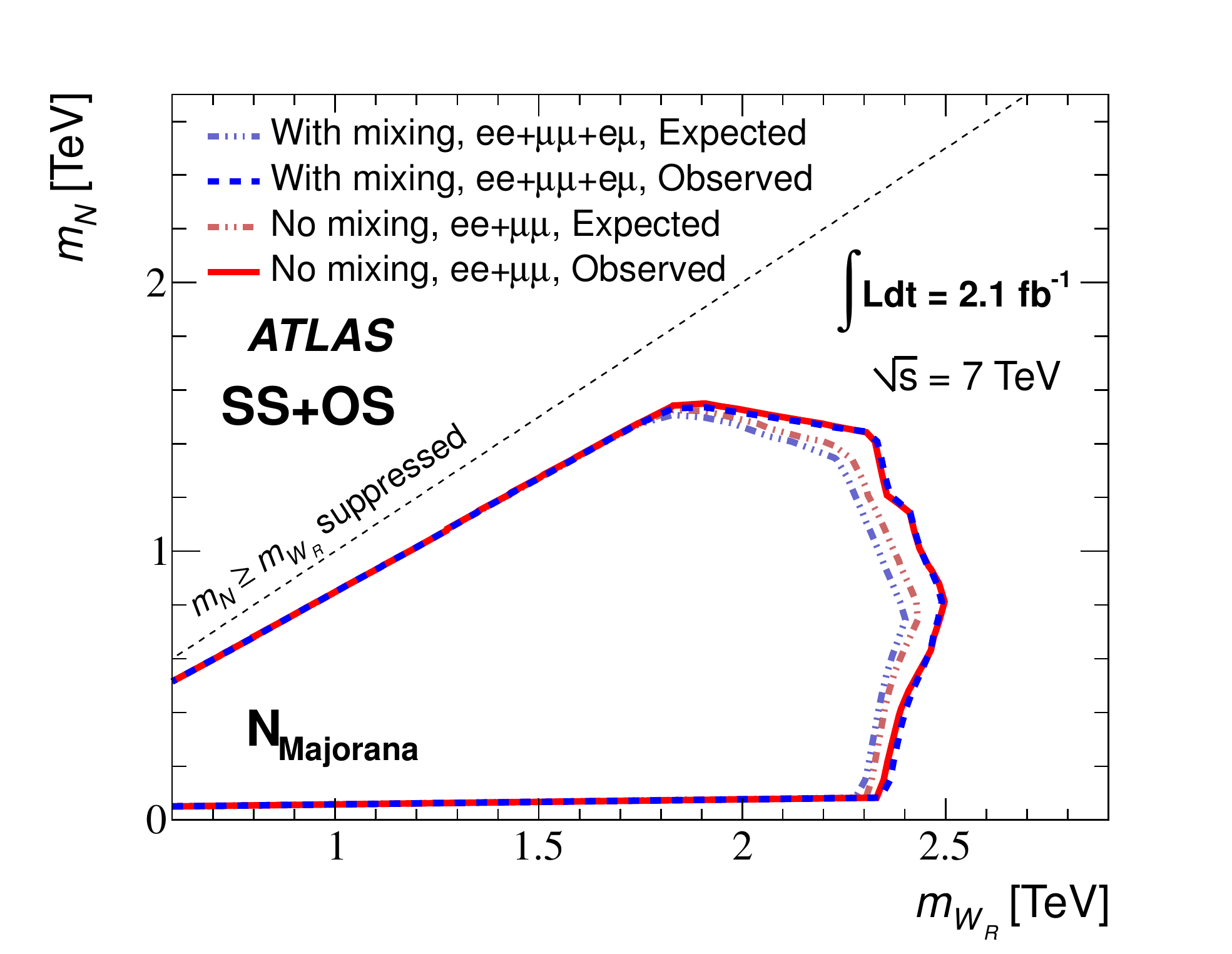}
\caption[*]{The 95\% confidence level exclusion region in the ($M_{W_R}$ ,$M_N$ ) plane for the electron channel from the CMS experiment (left) and for electron and muon channels combined from the ATLAS experiment (right).}
 \label{fig:NWR}
\end{figure}

The CMS and ATLAS experiments performed another search for a heavy neutrino and a right-handed $W_R$ boson~\cite{CMSNWR,ATLASNWR}. The masses of both the heavy neutrino and the $W_R$ boson are reconstructed and a template fit is applied to obtain the upper limit on the heavy neutrino and the $W_R$ boson masses as shown in Figure~\ref{fig:NWR}. 
 
\section{Searches for fourth-generation quarks}
\label{sect:TB}
The existence of a fourth generation of quarks would be another simple extension of the standard model. The CMS and ATLAS collaborations performed many searches for fourth-generation quarks, resulting in stringent constraints on the allowed parameter space. Two different types of models are distinguished. A first set of models predicts the existence of sequential fourth-generation quarks, while the second set of models predict the existence of vector-like fourth-generation quarks. The left- and right-handed components of the vector-like fourth-generation quarks transform in the same way under the weak force.

\begin{figure}[!thb]
  \centering
  \includegraphics[width=0.49 \textwidth]{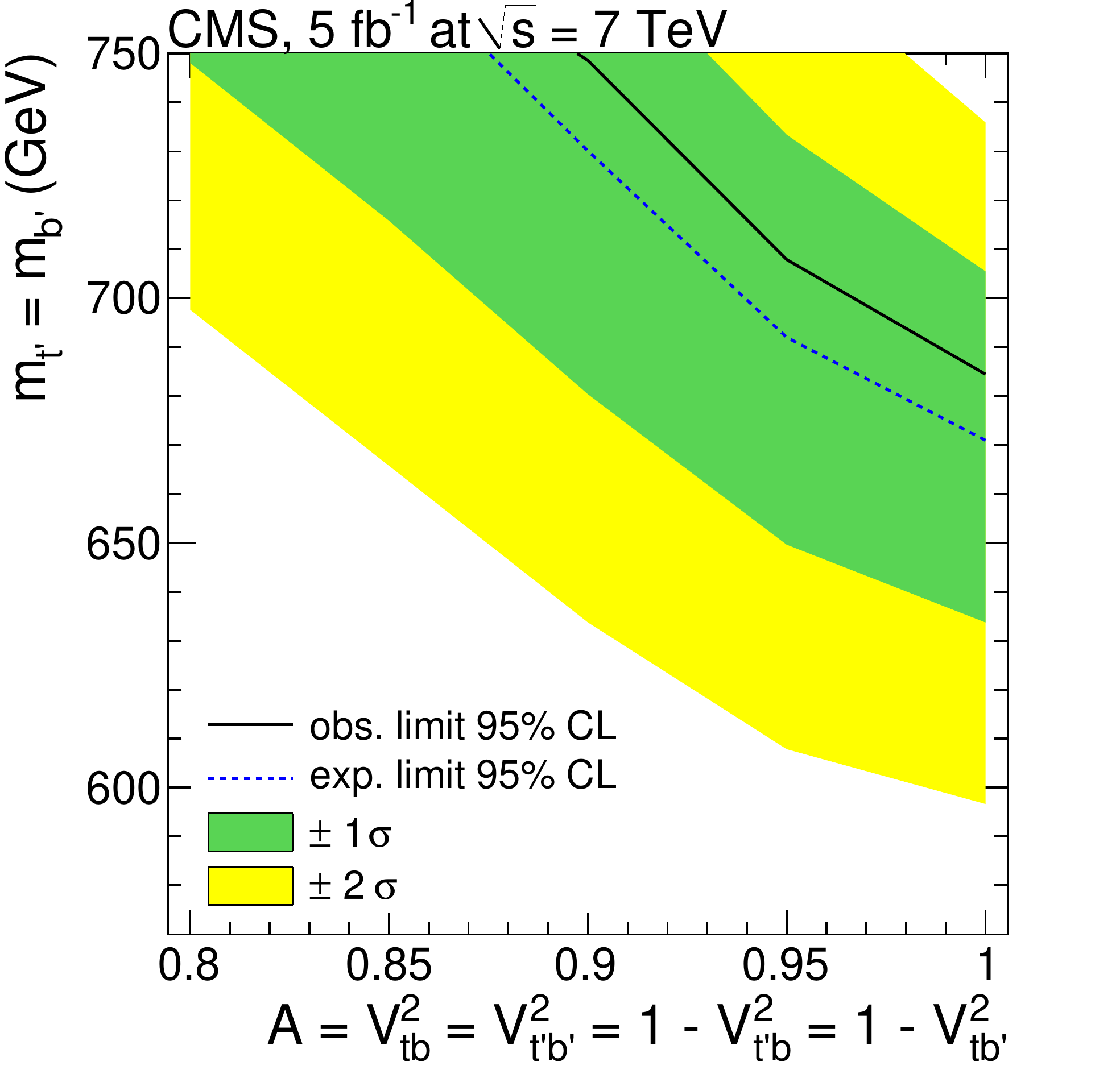}
  \includegraphics[width=0.49 \textwidth]{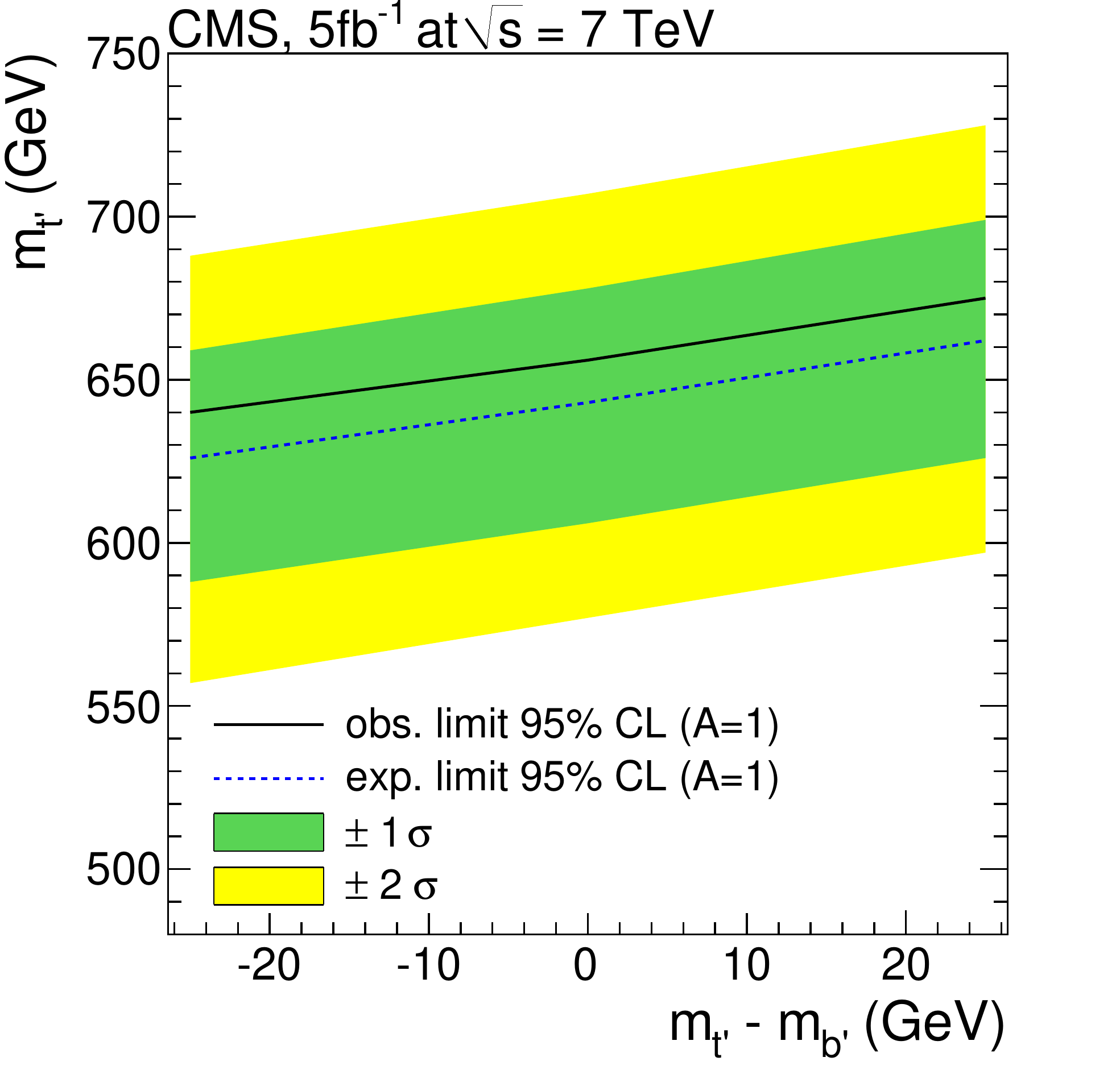}
\caption[*]{Left: The exclusion limit on $m_t' = m_b'$ as a function of the $V^{4 \times 4}_{CKM}$ parameter $A$. The
parameter values below the solid line are excluded at 95\% CL. The slope indicates the sensitivity of the analysis to the $t'b$ and $tb'$ processes. Right: For $A\sim1$, the exclusion limit on $m_{t'}$ versus $m_{t'} - m_{b'}$ is shown. The exclusion limit is calculated for mass differences up to 25 GeV.}
  \label{fig:CMScombinedTB}
\end{figure}

The CMS experiment developed a novel strategy for a combined search for sequential fourth-generation quarks of the up- and down-type in decay channels with at least one isolated muon or electron~\cite{CMScombinedTB}. A simple model, with a single parameter $A= |V_{tb}|^2 = |V_{t'b'}|^2$ is assumed for the extended CKM matrix, $V^{4 \times 4}_{CKM}$. Within this model, mixing is allowed only between the third and the fourth generations. Observables are constructed in each of the subsamples and used to differentiate between the standard model background and the processes with fourth-generation quarks. With this strategy the search for singly and pair-produced $t'$ and $b'$ quarks has been combined in a coherent way into a single analysis. 
Figure~\ref{fig:CMScombinedTB} shows the obtained model-dependent limits on the mass of the fourth-generation quarks and the relevant CKM matrix elements. The existence of mass-degenerate fourth-generation quarks with masses below 685 GeV is excluded at 95\% CL for minimal off-diagonal mixing between the third- and the fourth-generation quarks. With a mass difference of 25 GeV between the quark masses, the obtained limit on the masses of the fourth-generation quarks shifts by about $\pm$ 20 GeV. 

A search is performed by the CMS experiment for pair-produced vector-like down-type fourth-generation quarks, $B$, decaying to a $b$ quark and a $Z$ boson~\cite{CMS_VL_B}. The $Z$ boson is assumed to decay into a dielectron or dimuon pair. Events with two dilepton pairs of the same flavor and opposite charge and with an invariant mass between 60 and 120 GeV are selected. At least one jet should be identified as a $b$ jet with a high transverse momentum. The invariant mass of the $B$ quark is reconstructed by combining the $Z$ boson and $b$ jet that result in a reconstructed $B$ quark with the highest transverse momentum. A template fit is performed to obtain the upper limit on the production cross section times branching ratio as a function of the $B$ quark mass. The existence of a vector-like fourth-generation quark with mass below 550 GeV is excluded at the 95\% CL.

Some models predict the existence of a heavy partner of the top quark with a charge of 5/3, denoted as $T^{5/3}$. A search for this $T^{5/3}$ quark is developed by the CMS experiment assuming a branching fraction of 100\% for the decay of the $T^{5/3}$ quark to $tW$~\cite{CMS_T53}. After a dedicated event selection, the event yields are used to determine upper limits on the production cross section times branching fraction as a function of the fourth-generation quark mass. The existence of a $T^{5/3}$ quark with a mass below 645 GeV is excluded at the 95\% CL.

\begin{figure}[!thb]
  \centering
  \includegraphics[width=0.49 \textwidth]{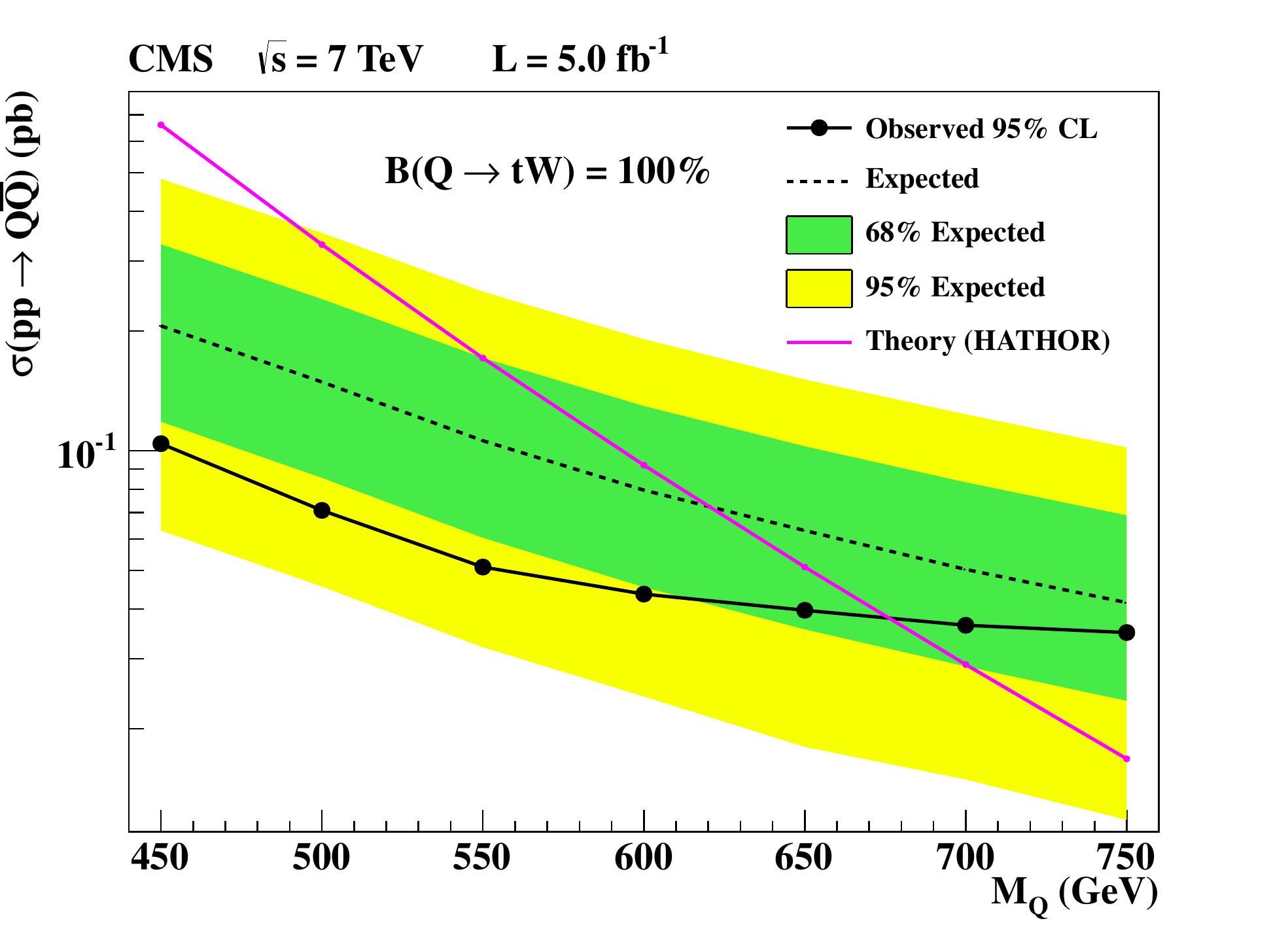}
  \includegraphics[width=0.49 \textwidth]{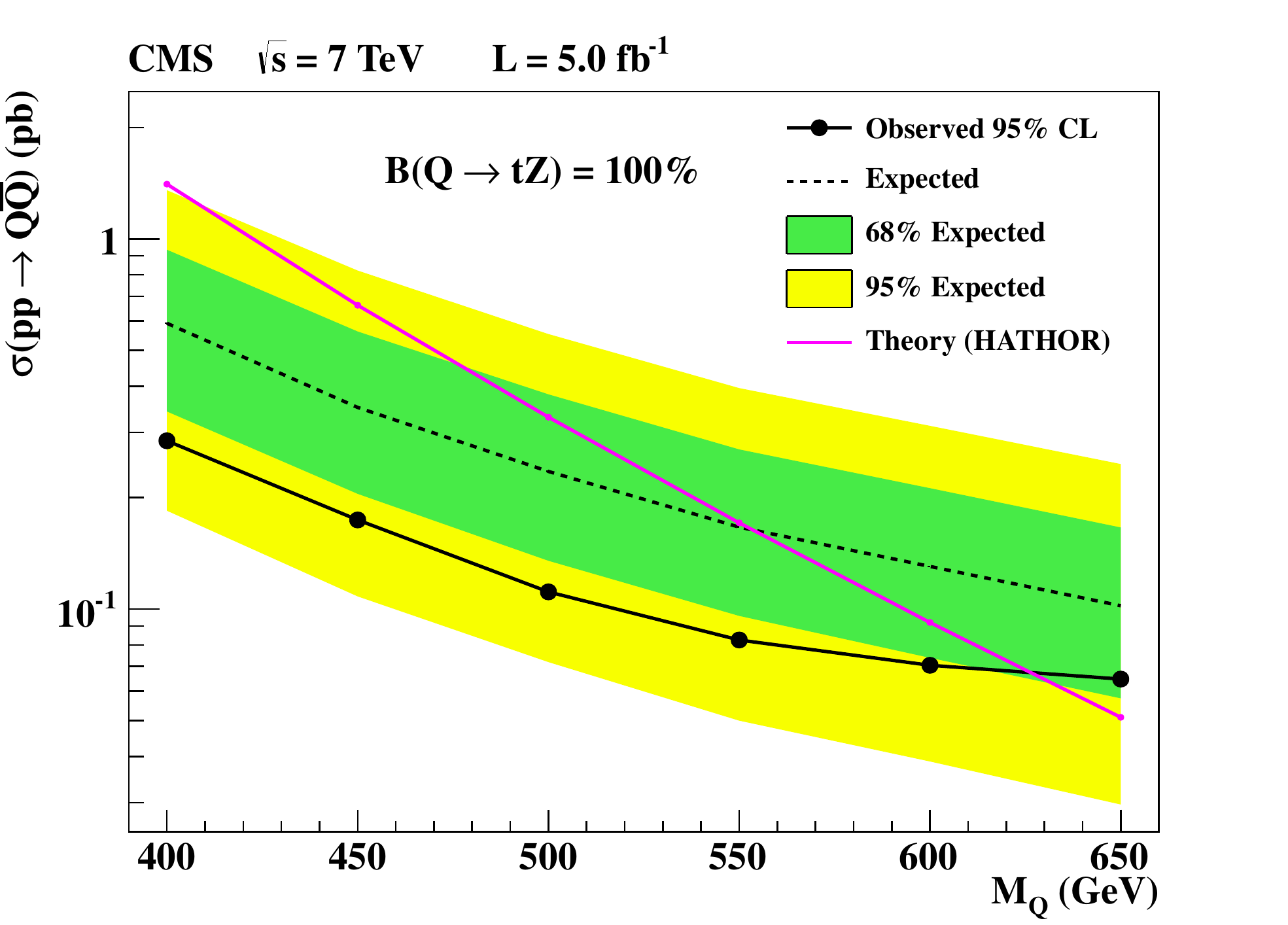}
\caption[*]{The left (right) plot shows the 95\% CL upper limit as a function of the $Q$ quark mass for a down-type sequential (up-type vector-like) heavy quark decaying exclusively to $tW$ ($tZ$).}
  \label{fig:B2G12004}
\end{figure}

The CMS experiment performed another search in the final state with one electron and muon and at least four jets~\cite{CMS_combined}. The distribution of the scalar sum of the transverse momenta of the reconstructed objects in the final state is obtained in different bins of the jet multiplicity and used in a template fit to obtain the upper limit on the production cross section times branching fraction. The resulting upper limit shown in Figure~\ref{fig:B2G12004} is obtained for two different models. The existence of a down-type sequential (up-type vector-like) fourth-generation quark with a mass below 675 (625) GeV is excluded at the 95\% CL.

\section{Summary}
\label{sect:summary}

We have summarized the most recent and the most stringent results from searches for new fermions and new bosons obtained by the ATLAS and CMS collaborations. Many more searches for new particles have been performed, but were not covered here. An example is the search for a resonance in the $t\bar{t}$ invariant mass distribution performed by the ATLAS and CMS collaborations, resulting in the exclusion of the existence of a $Z'$ boson decaying into $\rightarrow t\bar{t}$ with a mass below 1.55 TeV at the 95\% CL~\cite{CMS_ttbar1,ATLAS_ttbar1,CMS_ttbar2}. In Table~\ref{tab:overview} an overview is provided of the results from the searches discussed in the previous sections.
\begin{table}[hbtp]
\begin{center}
\caption{Overview of the limits on the masses of potential new fermions and bosons.}
\label{tab:overview}
\begin{tabular}{lc}
\hline
\hline
Search & 95\% CL limit\\
\hline
$Z' \rightarrow ee/\mu\mu$ &  $m_{Z'}> 2.6$ TeV \\
$Z' \rightarrow \tau\tau$ & $m_{Z'}> 1.4$ TeV  \\
$Z' \rightarrow qq$ &  $m_{Z'}> 1.6$ TeV \\
%$Z' \rightarrow bb$ & $m_{Z'}> 1.5$ TeV   \\
$Z' \rightarrow tt$ & $m_{Z'}> 1.55$ TeV  \\
\hline
$W' \rightarrow l\nu$ & $m_{W'}> 2.85$ TeV  \\
$W' \rightarrow tq$ & $m_{W'}> 1.1$ TeV  \\
\hline
$N \rightarrow W_Rl$ & $|V_{\mu N}|^2 < 0.07$ and $|V_eN|^2 < 0.22$ for $m_N=90$ GeV   \\
\hline
$t' \rightarrow bW$ and $b' \rightarrow tW$ & $m_{t'}=m_{b'}> 685$ GeV for $A=1$  \\
$B \rightarrow bZ$  &  $m_{B}> 550$ GeV \\
$T \rightarrow tZ$  &  $m_{T}> 625$ GeV \\
$T^{5/3} \rightarrow tW$  &  $m_{T^{5/3}}> 645$ GeV \\
\hline
\hline
\end{tabular}
\end{center}
\end{table}
Some theories beyond the standard model are already heavily constrained or ruled out. Furthermore, both the ATLAS and CMS collaborations continue to search for new fermions and bosons using the 8 TeV proton collisions. These efforts will result in more stringent limits or perhaps even a discovery of a new exotic particle.

%\section*{Acknowledgements} 

\end{document}